\documentclass[journal]{IEEEtran}
\usepackage{amsmath,amssymb,amsfonts,amsthm}
\newtheorem{theorem}{Theorem}
\newtheorem{definition}{Definition}

\usepackage{graphicx}
\usepackage{tikz}
\usetikzlibrary{decorations.pathreplacing,positioning,arrows.meta,calc,matrix}
\definecolor{tpclr}{RGB}{192,57,43}
\definecolor{epclr}{RGB}{41,98,168}
\usepackage{booktabs}
\usepackage{array}
\usepackage{multirow}
\usepackage{float}
\usepackage{url}
\usepackage[hidelinks]{hyperref}

\renewcommand{\thesection}{\arabic{section}}
\renewcommand{\thesubsection}{\thesection.\arabic{subsection}}
\renewcommand{\thetable}{\arabic{table}}
\renewcommand{\thefigure}{\arabic{figure}}
\makeatletter
\renewcommand{\thesectiondis}{\thesection}
\renewcommand{\thesubsectiondis}{\thesection.\arabic{subsection}}
\makeatother

\begin{document}

\title{Minimizing Optical Circuit Reconfigurations for Concurrent Collectives}
\author{
Xinchi Han\textsuperscript{1}\textsuperscript{*}, Weihao Jiang\textsuperscript{1}\textsuperscript{*}, Xuanyu Wang\textsuperscript{2}, Zhuoran Liu\textsuperscript{1}, Ximeng Liu\textsuperscript{1}, Haoyuan Liu\textsuperscript{1}, Kairui Zhou\textsuperscript{1}, Jiachen Wang\textsuperscript{3}, Yike Liu\textsuperscript{4}, Peirui Cao\textsuperscript{5}, Xun Nie\textsuperscript{6}, Xinbing Wang\textsuperscript{1}, and Shizhen Zhao\textsuperscript{1}\thanks{Shizhen Zhao is the corresponding author.}\\[2pt]
\textsuperscript{1}Shanghai Jiao Tong University, Shanghai, China\\
Emails: \{hanxinchi, weihao.jiang, cocopromenade-9, liuximeng, cse-lhy, kairuizhou, xwang8, shizhenzhao\}@sjtu.edu.cn\\
\textsuperscript{2}University of Electronic Science and Technology of China, Chengdu, China\\
Email: 2023010801006@std.uestc.edu.cn\\
\textsuperscript{3}Tsinghua University, Beijing, China\\
Email: wangjc22@mails.tsinghua.edu.cn\\
\textsuperscript{4}Xi'an Jiaotong University, Xi'an, China\\
Email: cn-lyk@stu.xjtu.edu.cn\\
\textsuperscript{5}State Key Laboratory of Novel Software Technology, Nanjing University, Nanjing, China\\
Email: caopeirui@nju.edu.cn\\
\textsuperscript{6}Xidian University, Xi'an, China\\
Email: xunnie@stu.xidian.edu.cn
\thanks{\textsuperscript{*}The first two authors contributed equally.}
}
\maketitle

\begin{abstract}
Optical circuit switches (OCSes) provide high-bandwidth point-to-point circuits in GPU clusters, but each optical port attaches to only one peer at a time, and reconfiguring a connection interrupts transmission. In large-model training, these ports are shared by many communication domains, each running its own collective instance, called a CCL kernel; often, several of these kernels execute the same collective at the same time. Because each step of the collective can reach a different peer, domains advancing through the steps in the same order may require a reconfiguration at every step.

We introduce the mega CCL kernel to schedule these kernels jointly. When a collective permits its steps to be reordered, different domains can take different legal steps in the same round, letting consecutive rounds share the same circuits and reducing the number of reconfigurations. Because the kernels are interchangeable, we need not decide the steps of every kernel separately. Instead, in each round we count how many kernels run each pattern, and then use bipartite matching to turn those counts into a concrete schedule for each kernel. We call this the count model, as opposed to the direct formulation that decides each kernel separately, and prove that the two give the same configuration sequences and the same optimum while the count model is substantially easier to solve.

On a fully loaded 64-supernode, 64-domain scheduling instance, joint scheduling cuts pairwise-AlltoAll reconfigurations from 62 to 3 and HD AllReduce from 10 to 2. In synthetic training replays parameterized by eight real models, it reduces iteration time by up to about 34\%; on a 64-GPU test case, the count model solves over six orders of magnitude faster than the direct formulation.
\end{abstract}

\section{Introduction}
\label{sec:intro}

Optical circuit switching (OCS) has been deployed at scale in large-model training clusters, most notably in Google's TPU v4~\cite{tpuv4}. An OCS connects its ports as point-to-point pairs, so each port serves at most one peer at a time, and changing that peer requires a reconfiguration that interrupts transmission for nanoseconds to milliseconds. A training collective often contacts several peers in one call, and switching among these peers can interrupt transmission repeatedly and waste bandwidth, and this paper studies how to minimize such interruptions under a fixed port budget.

A collective communication library (CCL), such as NCCL, runs a collective on a group of GPUs, which we call a communication domain. The collective instance executed on a domain is a CCL kernel. Large-model training combines tensor, data, expert, and pipeline parallelism, as in Megatron-LM~\cite{megatron}, which divides the GPUs into many domains sharing one set of optical ports. Such a deployment is highly regular: several domains of one type run the same collective at the same time on the same supernodes, with each domain using the same number of GPUs on every supernode. Many collectives contact different peers in different steps; pairwise AlltoAll, for example, exchanges data with each peer in turn. If all domains advance through these steps in the same order, they all request the same kind of connection in one round and switch together in the next, so the circuits may need reconfiguring at every step.

Our key observation is that the concurrent kernels of one collective do not have to perform the same step in the same round. Many collectives allow their steps to be reordered without changing the result, so we can assign different legal steps to different kernels in a round. One configuration can then carry all the links requested in a round and serve multiple consecutive rounds. The schedule can be arranged so that consecutive rounds keep the same set of optical circuits, assigning those circuits to different kernels as the rounds advance. The OCS then holds one configuration across these rounds and does not reconfigure, so the total number of reconfigurations drops sharply, while every kernel still completes the same steps with the same result.

We call this joint schedule a mega CCL kernel. It groups the independent kernels of one collective into a single large kernel and schedules their shared ports and bandwidth together. Two problems must be solved to build it. First, such a schedule requires a collective whose steps are interchangeable within a phase and whose circuit demands are consistent across domains. We characterize these conditions and show how to convert other collectives into such templates, including layered binary trees and irregular AlltoAllv. Second, choosing the connections of every kernel in every round on its own is slow and largely redundant. Since the kernels are interchangeable, we need not tell them apart at first. In each round we decide what links all the kernels need together, and then use bipartite matching to turn that decision into a concrete schedule for each kernel. We call this the count model, in contrast to the direct formulation, which is the joint formulation that keeps an explicit choice variable for every domain and round. We prove that the count model recovers every kernel's schedule and reaches exactly the same optimum, yet is much easier to solve.

This paper makes four contributions.
\begin{itemize}
\item We formulate the joint scheduling problem for the concurrent CCL kernels of one collective. A mega CCL kernel chooses each kernel's step order so that the shared OCS is reconfigured as few times as possible.
\item We give sufficient conditions under which a collective admits such a schedule, namely interchangeable steps and consistent circuit demands across kernels. We also show how to convert collectives that do not meet these conditions, including layered binary trees and irregular AlltoAllv.
\item We give the count model and use bipartite matching to prove that it recovers a legal schedule for every kernel with the same optimum as the direct formulation. For instances too large to solve exactly, we give feasible constructions that upper-bound the number of intervals.
\item On a fully loaded 64-supernode scheduling instance, the mega CCL kernel cuts pairwise-AlltoAll reconfigurations from 62 to 3 and HD AllReduce from 10 to 2. Across eight synthetic training replays it reduces iteration time by up to about 34\%, and on the hardest small instances the count model solves several orders of magnitude faster than the direct formulation.
\end{itemize}

\section{Related Work}
\label{sec:related}

\noindent\textbf{Keeping the traffic static.}
SCCL, TACCL, and TE-CCL synthesize the communication steps of a collective for a given topology~\cite{scccl,taccl,te_ccl}, while other work designs direct-connect topologies for specific collectives~\cite{zhao,basu}. Jeaugey's Parallel Aggregated Trees (PAT) gives a logarithmic AllGather and ReduceScatter algorithm on any number of ranks~\cite{pat}, and FAST balances skewed AlltoAllv traffic inside a server and schedules it as one-to-one transfers on an electrical network~\cite{fast}. SyCCL decomposes a collective into symmetric sub-demands and reuses the sub-schedules, and OptCCL synthesizes optimal collective algorithms on a fixed topology and jointly schedules several concurrent collectives that share network resources~\cite{syccl,optccl}. These works keep the underlying wiring static: they route and schedule traffic on a given network rather than change the discrete optical connections between rounds. The mega CCL kernel instead minimizes the number of OCS reconfiguration intervals over a fixed family of legal steps, and it is orthogonal to routing and synthesis algorithms that can produce the templates assumed here, and can be combined with such algorithms.

\noindent\textbf{Carrying only part of the traffic on the OCS.}
Helios, Opus, RailX, and MixNet approach the problem through network architecture: they place only the predictable traffic on optical circuits and leave the rest to electrical switches, or reconfigure the OCS only in the gaps between compute phases~\cite{helios,opus,railx,mixnet}. c-Through keeps a part-time optical fabric and lets the end hosts estimate traffic demand, Solstice schedules hybrid circuit and packet networks with a near-optimal heuristic, and a traffic-matrix scheduler drives microsecond-scale circuit switching, all coordinating the two fabrics under switching delay~\cite{cthrough,solstice,tms}. This requires the traffic to be cleanly separable, the parallel phases to be non-overlapping in time, or the traffic to have strong locality.

\noindent\textbf{Reconfigurable optical datacenter networks.}
RotorNet rotates the circuit matchings on a fixed cadence and decouples switch configuration from traffic, which removes demand estimation and admits a decentralized control plane~\cite{rotornet}. Sirius uses fast tunable lasers and passive gratings to present one high-radix optical switch with nanosecond reconfiguration~\cite{sirius}. OSA builds a flexible optical switching architecture that adapts its topology to the workload~\cite{osa}. Jupiter Evolving and Mission Apollo report Google's production use of MEMS optical switches and software-defined networking to transform its datacenter fabrics at scale~\cite{jupiter,apollo}. These systems engineer the whole network topology and its reconfiguration cadence, rather than count the independent legal step orders of the interchangeable kernels of one collective, which is the problem this paper addresses.

\noindent\textbf{Scheduling reconfiguration for a single kernel.}
Harvest and To Reconfigure or Not arrange the changing connections of one kernel, weighing reconfiguration time against the congestion and latency of indirect data paths~\cite{harvest,toreconf}. Bridge keeps a newly established connection in use across later steps~\cite{bridge}, SWOT runs reconfiguration in parallel with transmission and hides the switching time~\cite{swot}, and LACE decides at compile time which adjacent segments share a connection~\cite{lace}. All of them work along the time dimension of a single kernel and keep the relative step order across kernels unchanged, so they do not let different kernels take different steps in a round. When a kernel is highly dispersed and each GPU has only one uplink, there is little room to schedule that kernel in time. Our cross-domain scheduling works along a different dimension and can coexist with these methods.

\noindent\textbf{Redesigning collective libraries for reconfigurable optical networks.}
RECCL builds the reconfigurability of the optical network directly into the collective library, redesigning the communication relations inside a collective with a network cost model so that they fit a reconfigurable topology, and it targets reconfiguration on the order of hundreds of milliseconds~\cite{reccl}. Photonic PCCL adapts the network topology round by round to the communication pattern of a single collective and relies on an optimization framework to decide when to reconfigure~\cite{photonic_pccl}. Both center on reconfiguration within one collective and do not fuse the independent collectives running at the same time into one scheduling unit, so they schedule each collective on its own. The mega CCL kernel adds exactly this missing layer, without conflicting with the optimizations inside a single collective, and the two can be combined.

\noindent\textbf{Other related work.}
Chronos builds its network with crossbar circuit switches and stitches the local connections of concurrently running groups at the same call position into one global configuration, prescheduled in the inherent order; it does not reorder the steps of the groups or seek the minimum number of switches under independent step choices across domains~\cite{chronos}. Tang et al. group different jobs in an OCS data center, interleave computation and communication, and design the inter-rack optical topology stage by stage, with the job as the scheduling unit~\cite{tang}. Symphony coordinates multiple tenants online on an electrical network and overlaps steps that use different links to ease bandwidth contention~\cite{symphony}. These systems coordinate distinct, distinguishable entities, such as tenants, jobs, or call positions, and schedule them in their given order or overlap steps that use separate links. They do not exploit the freedom of the interchangeable domains of one collective to choose separate legal step orders and then count the shared OCS per round, which is the distinction this paper develops.

The mega CCL kernel starts from a different premise. Spatially, the communication domains in a large-model deployment are distributed across the supernodes in a regular, homogeneous way and share the same set of optical ports. At the operator level, the steps of a collective can be interchanged and reordered when they have no dependencies, without changing the result. This paper combines the two: interchangeable kernels take different legal steps in the same round, and the count model records these arrangements rather than deciding each kernel separately, which greatly lowers the solving cost.

\section{Motivation and Problem Formulation}
\label{sec:motivation}

This section first describes how the concurrent CCL kernels are placed in large-model training and where OCS reconfigurations come from, and it sets up the system model. It then explains the freedom of kernels running the same collective to reorder their steps and the benefit of doing so, and formalizes the arrangement as a solvable optimization problem. Finally, it states the two challenges that must be addressed to turn this arrangement into a mega CCL kernel.

\subsection{Spatial Placement of Communication Domains}
We consider a cluster of supernodes. A supernode is a rack of GPUs joined by a high-bandwidth local fabric, such as NVLink/NVSwitch, UB, or IB, and it connects to the OCS through only a limited number of optical uplinks. Within a supernode, each GPU can address any uplink through memory semantics, so it can be reassigned to a different uplink at the boundary between communication steps. The only constraint is the total number of uplinks a supernode can use at any time.

Training frameworks combine tensor, data, expert, and pipeline parallelism. Before training starts, they partition all the GPUs into groups, one group for each parallel dimension. Each group is a communication domain whose GPUs run collectives together, and the collective instance running on a domain is a CCL kernel. A collective reaches only the GPUs in its own domain. A GPU can belong to one domain in each of several parallel dimensions and take part in their communication at different times in an iteration.

A single GPU can belong to several domains, because training combines several forms of parallelism. Each form of parallelism can be viewed as an independent axis along which the GPUs are indexed and grouped. Domains on different axes are independent and are called orthogonal, as with the EP and PP domains that one GPU can join; the several domains of one type instead lie on one axis and are not orthogonal to each other. If a GPU joins two orthogonal domains at once, it needs several uplinks toward different supernodes, and the limited port count may not cover them. Work such as Opus and RailX lets a GPU take part in different domains at different times and hide the OCS switching cost inside computation. We follow this time-staggered arrangement so each GPU joins only one inter-supernode domain at a time. Our technical report explains how to design the mega CCL kernel when orthogonal domains may communicate at the same time.

\begin{figure}[t]
\centering
\includegraphics[width=\columnwidth]{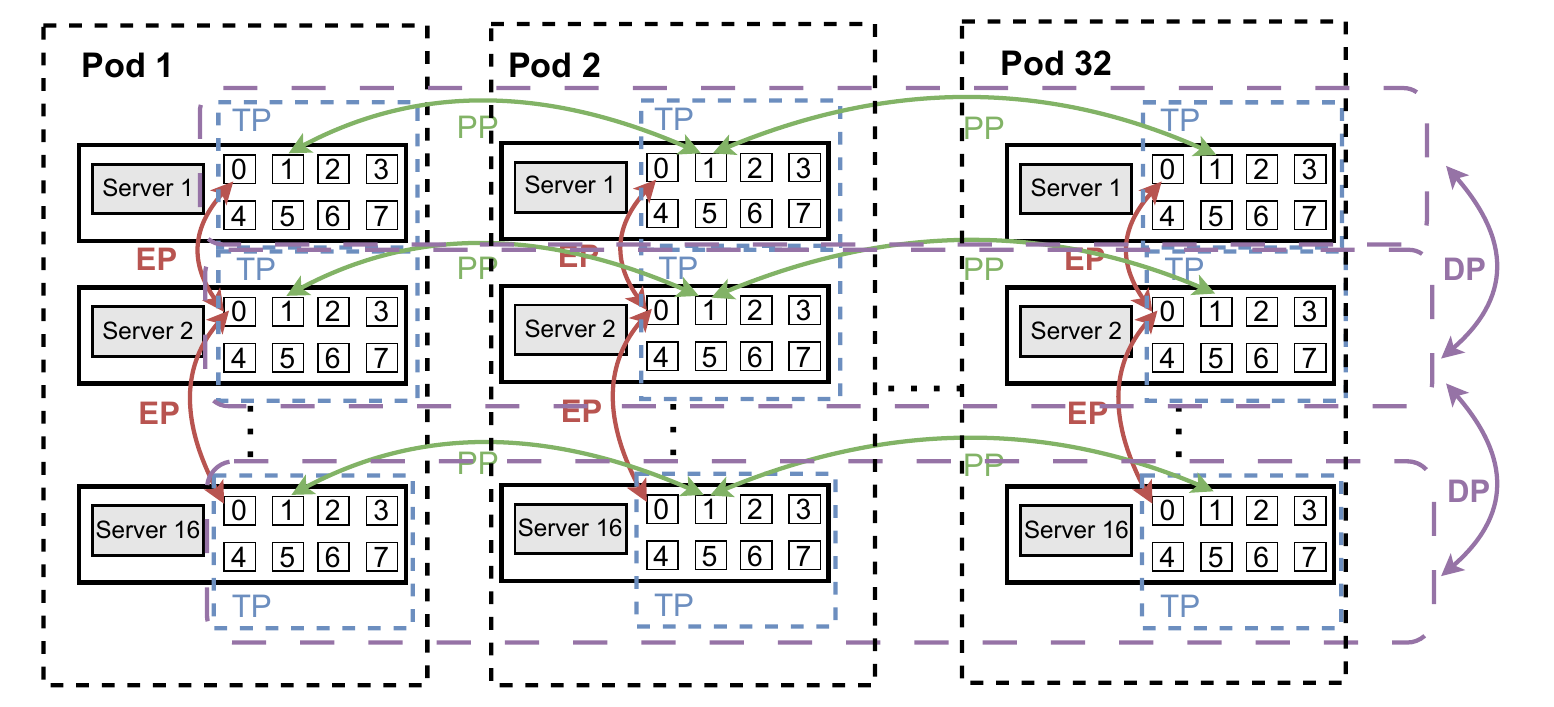}
\caption{One placement of TP, EP, PP, and DP domains. Each pod contains several supernodes with locally numbered GPUs. TP groups stay within a supernode (blue dashed boxes). EP connects equal local indices across supernodes (red arrows). PP connects stages across pods (green arcs), and DP replicates the outer groups (purple arrows). In this example several PP groups use the OCS concurrently. Other placements can carry DP, PP, or EP over the OCS. Domains of the same type use equal numbers of GPUs on the same supernodes.}
\label{fig:domains}
\end{figure}

These domains are organized in spatial layers. A single TP step moves a large amount of data and usually stays inside one supernode over the local fabric, without using any optical port. DP, EP, and PP, which are orthogonal to TP, often cross supernodes and rely on the OCS to reconfigure the light paths. Take $\text{TP}=4$ and $\text{EP}=32$. The GPUs with the same index inside each TP group form one EP domain: the index-0 GPUs across all TP groups make up one EP domain, the index-1 GPUs make up another, and so on. The memory, local-fabric size, and GPU count of one supernode are all bounded, which limits the TP size a single supernode can hold. As a result, when DP, EP, or PP crosses supernodes, the resource use is highly regular: the several domains of one type occupy the same set of supernodes with equal numbers of GPUs and uplinks. We rely on exactly this regular structure. Fig.~\ref{fig:domains} depicts the deployment at the three levels of pod, server, and GPU, and uses colored boxes and arrows to mark the GPUs and directions of the four types of domain, TP, EP, PP, and DP.

The OCS carries the inter-supernode traffic by setting up circuits between supernode uplinks. One circuit joins two ports and lets them exchange data in both directions, and all circuits present at one time together form a configuration. Each port belongs to at most one circuit. To change a connection, the ports must be rewired, which is a reconfiguration and takes time.

We focus on this question: when several CCL kernels execute the same collective, can they adopt different, yet all legal, orders so that as many steps as possible share one optical configuration? Because reconfiguration carries a cost, frequent switching pulls down throughput even when bandwidth is ample. We therefore aim to keep reconfigurations few, and then reuse the OCS logical topology until its bandwidth approaches that of an electrical cluster.

\subsection{System Model}
There are $M$ supernodes and $G$ concurrent domains. Each domain uses $N$ GPUs on every supernode, so a domain contains $M \times N$ GPUs in total and a supernode holds $G \times N$. The case $N=1$ is the simplest, where each domain contributes one GPU to every supernode. This equal placement is guaranteed by the spatial distribution described above, and it holds regardless of where TP is placed or whether the parallel dimensions are orthogonal.

Each supernode has $H$ optical uplinks, and the OCS can realize any inter-supernode circuit multigraph that fits these budgets, where the circuit multiplicities between supernode pairs define a configuration. The setting we focus on is fully provisioned: every GPU has one uplink, so $H=G \times N$, and all ports are occupied once the GPUs use inter-supernode circuits. Rebinding a GPU to another local uplink does not reconfigure the OCS. The scheduler only chooses among the step orders and rank mappings allowed by the template, with every logical inter-supernode transfer carried by a direct circuit; all optimality claims are relative to this deployment and template.

\subsection{How Domains Can Share a Configuration}
\label{sec:benefit}

Under the regular placement, when several CCL kernels run the same collective, they all have the same steps to complete, but these steps do not have to be taken in the same order. We exploit exactly this freedom by letting different CCL kernels execute different steps at the same time, which cuts the number of OCS switches. We first walk through the smallest example to make the mechanism clear.

Let $M=G=4$, $N=1$, and $H=4$. Four XOR AlltoAll domains span four supernodes, and each supernode has four optical uplinks. Within a domain, its four GPUs are numbered ranks $0,1,2,3$, with rank $i$ placed on supernode $i$. The XOR AlltoAll takes three steps, and each step pairs the four ranks into two pairs: first rank $0$ with $1$ and rank $2$ with $3$, then rank $0$ with $2$ and rank $1$ with $3$, and finally rank $0$ with $3$ and rank $1$ with $2$. Under this placement, each rank pair maps directly to a circuit between the corresponding supernodes, so these three steps translate into three fixed actions, or patterns:
\begin{itemize}
\item Pattern $a$ pairs supernodes 0--1 and 2--3.
\item Pattern $b$ pairs supernodes 0--2 and 1--3.
\item Pattern $c$ pairs supernodes 0--3 and 1--2.
\end{itemize}

To describe the example precisely, we separate two questions: what circuits a step needs, and when that step executes. We capture the first with a pattern and the second with a round.

\noindent\textbf{Patterns describe circuit demand.}
The collective defines a set of actions, each being a fixed logical connection that specifies which GPUs exchange data. When this logical connection is combined with the GPU placement, that is, how the $N$ GPUs and ranks map onto physical cards in each supernode, it becomes a concrete pattern stating which supernodes are connected and by how many circuits. A pattern is thus determined jointly by the collective and the placement: it describes only the connection demand and is not tied to particular ports or a particular kernel. For example, no matter which kernel runs pattern $a$, the supernode layer always builds the two circuits 0--1 and 2--3.

\noindent\textbf{Rounds specify when steps execute.}
All CCL kernels align to the same sequence of time windows, called rounds, and each kernel executes one pattern per round, though different kernels may execute different patterns. Here every kernel must run $a,b,c$ once, so the collective takes three rounds. Because the three pairwise exchanges have no data dependencies, each kernel is free to choose their order.

If all CCL kernels follow $a\to b\to c$, their aggregate demands are $4a$, $4b$, and $4c$. The three rounds require three configurations and two reconfigurations.

Table~\ref{tab:toy} gives another schedule. Kernels 0 through 3 follow $(a,b,c)$, $(a,c,b)$, $(b,a,c)$, and $(c,a,b)$, respectively. Each completes all three exchanges.

\begin{table}[!t]
\centering
\caption{Four CCL kernels execute three rounds. Each cell gives the pattern for one kernel in one round. The last row counts kernels per pattern. For example, $2a+b+c$ assigns two kernels to $a$ and one each to $b$ and $c$.}
\label{tab:toy}
\footnotesize
\setlength{\tabcolsep}{10pt}
\begin{tabular}{lccc}
\toprule
& Round 1 & Round 2 & Round 3 \\
\midrule
Kernel 0 & $a$ & $b$ & $c$ \\
Kernel 1 & $a$ & $c$ & $b$ \\
Kernel 2 & $b$ & $a$ & $c$ \\
Kernel 3 & $c$ & $a$ & $b$ \\
\midrule
Counts & $2a+b+c$ & $2a+b+c$ & $2b+2c$ \\
\bottomrule
\end{tabular}
\end{table}

The first two rounds have the same aggregate demand, $2a+b+c$. The specific kernels carrying each pattern change, but in both rounds two kernels run $a$ and one each run $b$ and $c$, so both need two circuits on 0--1 and two on 2--3 for the $2a$, plus one copy of every edge in $b$ and $c$. Since the OCS only fixes how many circuits connect each supernode pair and not which kernel uses them, local uplink reassignment lets both rounds share one optical configuration. Only the third round, whose demand is $2b+2c$, needs a change. The schedule therefore spans two configuration intervals with a single reconfiguration.

Both configurations use exactly four uplinks per supernode. Each kernel still executes three rounds and sends each message once. With equal windows and link rates and negligible local reassignment cost, the schedule saves one reconfiguration without increasing transmission time.

In short, a pattern states what connections one kernel needs, a round aligns all kernels to the same time window, and an OCS configuration is determined by the aggregate demand of every kernel in that round. By reordering the patterns so that consecutive rounds produce the same aggregate demand, the kernels can reuse one configuration across those rounds and reduce the number of reconfigurations.

\begin{figure}[!t]
\centering
\begin{tikzpicture}[font=\scriptsize,
  g/.style={circle, draw=black!55, fill=black!4, line width=0.4pt,
    minimum size=0.40cm, inner sep=0pt}]
\foreach \i in {0,...,7}{\node[g] (a\i) at (\i*0.78,0) {\i};}
\foreach \i in {0,...,7}{\node[g] (b\i) at (\i*0.78,-0.90) {\i};}
\foreach \i in {0,...,7}{\node[g] (c\i) at (\i*0.78,-1.80) {\i};}
\foreach \i/\j in {0/1,2/3,4/5,6/7}{\draw (a\i) to[bend left=45] (a\j);}
\foreach \i/\j in {0/2,4/6}{\draw (b\i) to[bend left=40] (b\j);}
\foreach \i/\j in {1/3,5/7}{\draw (b\i) to[bend right=40] (b\j);}
\foreach \i/\j in {0/4,2/6}{\draw (c\i) to[bend left=22] (c\j);}
\foreach \i/\j in {1/5,3/7}{\draw (c\i) to[bend right=22] (c\j);}
\foreach \i in {0,...,7}{\node[g] at (a\i) {\i};}
\foreach \i in {0,...,7}{\node[g] at (b\i) {\i};}
\foreach \i in {0,...,7}{\node[g] at (c\i) {\i};}
\node[anchor=east] at (-0.35,0) {Offset 1};
\node[anchor=east] at (-0.35,-0.90) {Offset 2};
\node[anchor=east] at (-0.35,-1.80) {Offset 4};
\node[anchor=east] at (-0.35,-2.95) {Kernel 0};
\node[anchor=east] at (-0.35,-3.57) {Kernel 1};
\node[anchor=south] at (0.72,-2.72) {RS};
\node[anchor=south] at (2.88,-2.72) {AG};
\foreach \i/\s in {0/1,1/2,2/4}{\node[fill=orange!18, draw=black!35, line width=0.4pt,
  minimum width=0.58cm, minimum height=0.34cm, inner sep=0pt] at (\i*0.72,-2.95) {\s};}
\foreach \i/\s in {3/4,4/2,5/1}{\node[fill=epclr!14, draw=black!35, line width=0.4pt,
  minimum width=0.58cm, minimum height=0.34cm, inner sep=0pt] at (\i*0.72,-2.95) {\s};}
\foreach \i/\s in {0/1,1/4,2/2}{\node[fill=orange!18, draw=black!35, line width=0.4pt,
  minimum width=0.58cm, minimum height=0.34cm, inner sep=0pt] at (\i*0.72,-3.57) {\s};}
\foreach \i/\s in {3/2,4/4,5/1}{\node[fill=epclr!14, draw=black!35, line width=0.4pt,
  minimum width=0.58cm, minimum height=0.34cm, inner sep=0pt] at (\i*0.72,-3.57) {\s};}
\draw[dashed, black!45] (1.80,-2.74) -- (1.80,-3.78);
\node[anchor=east] at (-0.35,0.50) {\textbf{(a)}};
\node[anchor=east] at (-0.35,-2.50) {\textbf{(b)}};
\end{tikzpicture}
\caption{HD AllReduce on eight ranks. (a) The three matchings have offsets 1, 2, and 4. (b) Kernel 0 uses $1\to2\to4$ for RS and $4\to2\to1$ for AG. Kernel 1 uses $1\to4\to2$ and $2\to4\to1$. Both schedules use a full matching in each of six rounds.}
\label{fig:hd_reorder}
\end{figure}
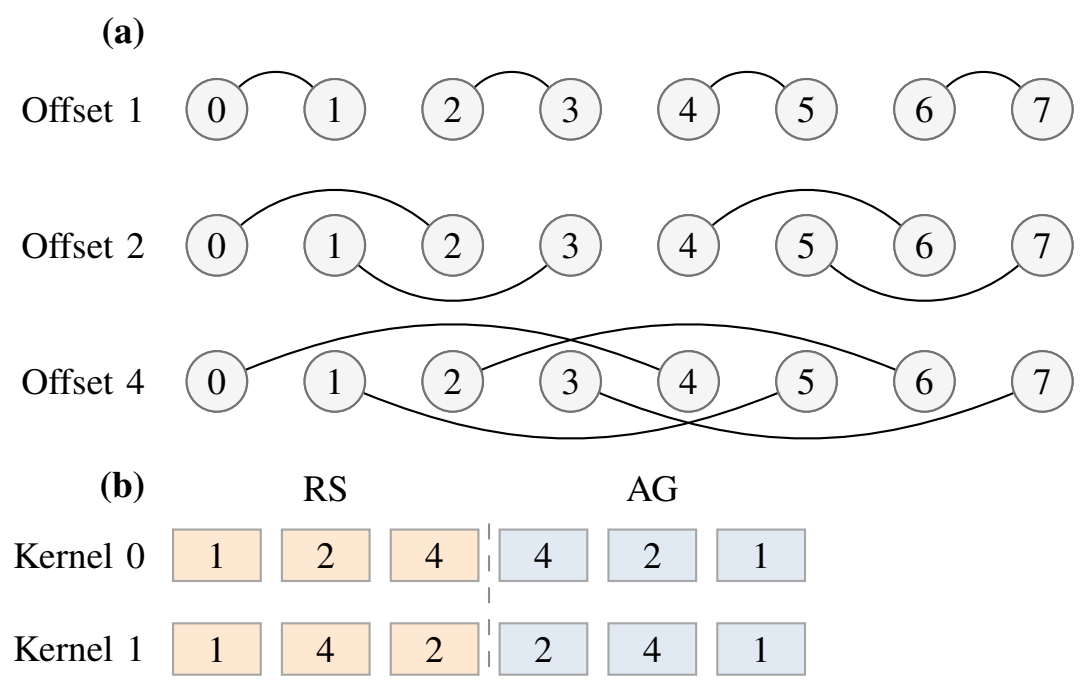

The AlltoAll example above works because all its steps are mutually independent. To see whether the same reordering freedom survives when steps carry dependencies, and where its limits are, consider HD AllReduce with $M=8$ and $N=1$. It consists of two phases, ReduceScatter (RS) and AllGather (AG). RS first leaves each rank with a distinct part of the reduced tensor, and AG then distributes those parts to all ranks, so AG must follow RS and the two phases cannot be swapped. Each phase visits three dimensions, pairing rank $r$ with $r\oplus1$, $r\oplus2$, and $r\oplus4$ (Fig.~\ref{fig:hd_reorder}a). Within either phase, a kernel can choose a dimension order and compile the corresponding block exchanges and reductions. Fig.~\ref{fig:hd_reorder}b shows two schedules that use the same matchings in different orders. Our technical report gives a compilation proof for arbitrary within-phase dimension orders.

The reordering seen in these examples relies on two conditions, one temporal and one spatial.

\textbf{Temporal condition: steps within a kernel can be reordered.}
When several steps have no data dependency between them, a kernel may execute them in any order and still obtain the same result, whereas steps that carry a dependency cannot be moved. For example, within either phase of HD the three dimensions can be visited in any order, provided the block exchanges are compiled for that order, but AG cannot be moved ahead of RS.

\textbf{Spatial condition: a pattern's demand is independent of which kernel runs it.}
No matter which kernel carries pattern $a$, at the supernode layer it always connects the same pair set, 0--1 and 2--3. The OCS only fixes how many circuits connect each supernode pair and never which kernel uses them, so the same pattern yields the same demand in every kernel, which is what lets the circuits of different kernels merge into a single configuration. The patterns $a,b,c$ are pairings at the GPU layer, and turning them into supernode-layer demand requires aggregating them by supernode: for each supernode pair, count the GPU edges that fall between them and call the resulting matrix $B_v$ for pattern $v$. When $N=1$, one GPU pairing contributes at most one inter-supernode circuit; when $N>1$, several pairings can aggregate onto the same supernode pair. Section~\ref{sec:problem-def} gives the formal definition.

\noindent\textbf{Remark:} Joint scheduling requires all kernels to share round boundaries. Real collectives are usually hierarchical: they first communicate inside each supernode and then move to inter-supernode communication, so the local completion point is a common boundary at which the kernels align. A kernel that enters inter-supernode communication without this local stage needs explicit synchronization instead. If every kernel follows a common timetable and finishes each round's work inside its window, one synchronization at the start keeps them aligned throughout. In practice, GPUs may communicate at different rates as their operating conditions vary, so synchronization between rounds is optional. Whether to synchronize is the same tradeoff studied in Section~\ref{sec:switch-speed}: a brief synchronization keeps the phases aligned and prevents drift, while skipping it avoids the extra delay. With large messages the cost is small and synchronization keeps the phases from drifting; with small messages it may add latency.

\begin{figure}[!t]
\centering
\resizebox{\columnwidth}{!}{%
\begin{tikzpicture}[font=\scriptsize,
  lane/.style={anchor=east},
  blk/.style={draw=black!50, line width=0.4pt, rounded corners=2pt,
    minimum height=0.44cm, inner sep=2pt},
  comm/.style={draw=blue!60, fill=blue!8, line width=0.4pt, rounded corners=2pt,
    minimum height=0.38cm, inner sep=2pt},
  inner/.style={draw=black!40, fill=black!8, line width=0.4pt, rounded corners=2pt,
    minimum height=0.38cm, inner sep=2pt},
  sw/.style={draw=orange!85!black, fill=orange!22, line width=0.4pt, rounded corners=2pt,
    minimum height=0.40cm, inner sep=2pt},
  cfg/.style={line width=0.8pt, draw=black!75}]
\node[lane] at (-0.15,1.5) {Phase};
\node[lane] at (-0.15,0.5) {Collective};
\node[lane] at (-0.15,-0.5) {Circuits};
\node[blk, fill=yellow!24, minimum width=2.0cm] at (1.3,1.5) {CP};
\node[text=black!60, font=\tiny] at (3.0,1.5) {Local compute};
\node[blk, fill=green!20, minimum width=2.6cm] at (5.0,1.5) {EP};
\node at (6.75,1.5) {$\cdots$};
\node[blk, fill=yellow!24, minimum width=1.6cm] at (8.0,1.5) {CP};
\node[comm, minimum width=1.0cm] at (1.1,0.5) {Ring};
\node[inner, minimum width=1.0cm] at (3.0,0.5) {Local ring};
\node[comm, minimum width=0.9cm] at (4.45,0.5) {A2A};
\node[comm, minimum width=0.9cm] at (5.55,0.5) {A2A};
\node[comm, minimum width=1.0cm] at (8.0,0.5) {Ring};
\draw[cfg] (0.3,-0.5) -- (2.3,-0.5);
\node[anchor=south] at (1.3,-0.46) {Config. A};
\draw[cfg] (3.7,-0.5) -- (6.3,-0.5);
\node[anchor=south] at (5.0,-0.46) {Config. B};
\draw[cfg] (7.2,-0.5) -- (8.8,-0.5);
\node[anchor=south] at (8.0,-0.46) {Config. A};
\node[sw, minimum width=1.0cm] at (3.0,-0.5) {Switch};
\node[sw, minimum width=0.7cm] at (6.75,-0.5) {Switch};
\draw[dashed, black!50, line width=0.4pt] (3.0,-0.95) -- (3.0,0.15);
\draw[dashed, black!50, line width=0.4pt] (6.75,-0.95) -- (6.75,0.15);
\draw[->, line width=0.7pt] (0,-1.4) -- (9.2,-1.4);
\node[anchor=west] at (9.25,-1.4) {Time};
\end{tikzpicture}}%
\caption{Reconfiguration at parallel-phase boundaries. Context parallelism (CP) runs Ring, and expert parallelism (EP) runs two AlltoAll calls (A2A). Local computation and an intra-supernode Ring occupy the intervening gap. The OCS can reconfigure in that gap (orange) if it is long enough.}
\label{fig:timing}
\end{figure}
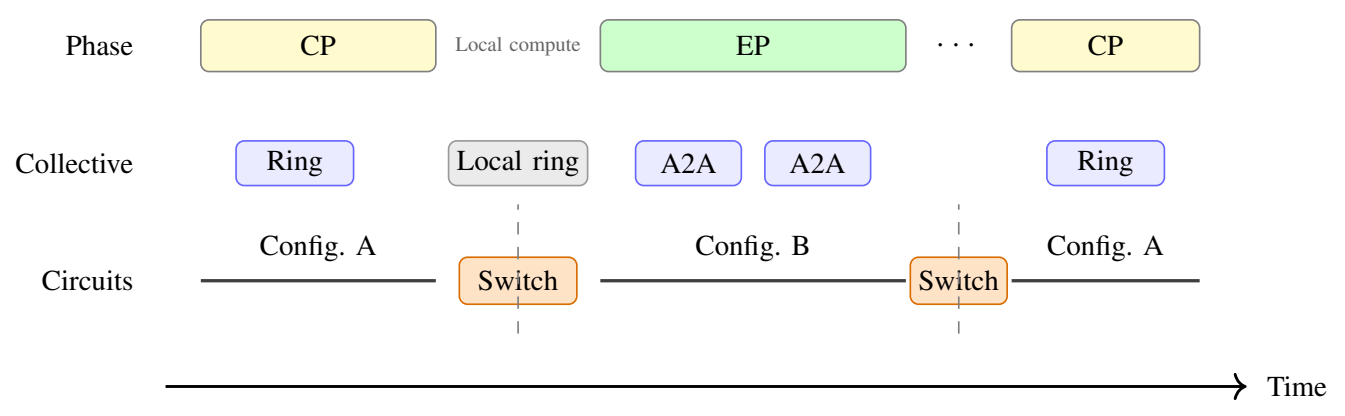

\subsection{Legal Executions and Homogeneous Templates}
\label{sec:problem-def}
We now express these two conditions in precise notation so that the scheduling problem can be stated formally.

A mega CCL kernel combines the $G$ CCL kernels that run the same collective and spans $V$ rounds, with each CCL kernel executing one pattern per round. Let $\mathcal V$ be the set of distinct patterns. A pattern specifies the actions of all $M \times N$ GPUs of one CCL kernel in one round. Patterns may repeat, and a local-only pattern occupies a round with zero inter-supernode demand.

Let $E_v$ be the circuit count for pattern $v$, a function on unordered GPU pairs: $E_v(\{a,a'\})=k$ means GPUs $a$ and $a'$ need $k$ dedicated full-duplex circuits, parallel edges when $k>1$, and $E_v(\{a,a'\})=0$ means they do not communicate. For distinct supernodes $u$ and $y$, aggregating over their GPUs gives the circuit demand
\begin{equation}
B_v(u,y)=\sum_{m=1}^{N}\sum_{n=1}^{N}E_v\!\left(\{(u,m),(y,n)\}\right),\quad u\ne y,
\label{eq:demand}
\end{equation}
The OCS connects supernodes with full-duplex circuits, so the demand of pattern $v$ is described by the symmetric matrix $B_v$, which is nonnegative and integer-valued with a zero diagonal. Each entry counts the full-duplex circuits that pattern $v$ requires between two supernodes. Local edges consume no optical ports, and when each GPU uses at most one inter-supernode circuit, $\sum_{y\ne u} B_v(u,y)\le N$. These demands can be added across independent scheduling units.

Within the mega CCL kernel considered here, all the CCL kernels share the same pattern multiplicities. We partition the $V$ rounds into consecutive phases $I_1,\ldots,I_P$. A CCL kernel may reorder the steps inside a phase, while the phases themselves stay in their original order.

Let $\mu_p(v)$ be the number of times each CCL kernel must execute pattern $v$ over the whole phase $p$. This phase-level total is fixed by the collective and is the same for every kernel, but it does not specify which pattern a kernel runs in any particular round, so the kernels can still carry different patterns in the same round by ordering these executions differently. Then
\begin{equation}
\sum_v\mu_p(v)=|I_p|,\qquad
\mu(v)=\sum_p\mu_p(v),\qquad \sum_v\mu(v)=V.
\end{equation}
Let $\Omega_p$ contain all permutations of that phase's multiset of steps.

We now define a homogeneous communication template.

\begin{definition}[Homogeneous communication template]
\label{def:homogeneous}
Suppose $G$ CCL kernels execute the same collective, each on a domain using $N$ GPUs per supernode, and they share a pattern set $\mathcal V$, phase boundaries $I_p$, and multiplicities $\mu_p$. The template has a \emph{homogeneous representation} if it satisfies the following properties.

\noindent\emph{1. Within-phase permutations preserve the result.}
For every
\begin{equation}
(\sigma_1,\ldots,\sigma_P)\in\Omega_1\times\cdots\times\Omega_P
\label{eq:phase-product}
\end{equation}
each CCL kernel can compile the phase orders $\sigma_1,\ldots,\sigma_P$ into an execution with the same collective result as the original order. Reductions assume an associative and commutative operator, and the template preserves phase boundaries.

\noindent\emph{2. Patterns have kernel-independent circuit demand.}
After applying each kernel's rank placement, pattern $v$ requires the same number of circuits between every pair of physical supernodes in every kernel; this common physical demand is the matrix $B_v$ from Equation~\eqref{eq:demand}. With a single port per GPU, each GPU's degree, counted with multiplicity $\sum_{a'\ne a}E_v(\{a,a'\})$, is at most one, and when all GPUs participate in such pairings, the edges form a perfect matching with $M \times N/2$ edges. The same aggregation applies to multiport patterns, rings, and patterns with idle GPUs.
\end{definition}

The phase permutations specify the pattern executed in each round. A CCL kernel assigned $v$ in round $s$ requires $B_v$ in that round, and the matrices $B_v$ and phase lists $\mu_p$ are inputs to the joint scheduler.

\subsection{Inputs, Constraints, and Outputs}
\label{sec:io}
Given hardware resources, domain placement, and a homogeneous communication template, the scheduler solves for the mega CCL kernel defined above. It assigns each CCL kernel one pattern in every round and sets the optical configuration that carries the aggregate demand in that round.

\noindent\textbf{Inputs.}
\begin{enumerate}
\item \textbf{Hardware.} There are $M$ supernodes, $G$ domains, $N$ GPUs per domain per supernode, and $H$ uplinks per supernode. Each domain contains $M \times N$ GPUs and each supernode contains $G \times N$. GPUs can be reassigned to local uplinks at round boundaries, and with one uplink per GPU, $H=G \times N$.
\item \textbf{Placement.} The input identifies the GPUs and supernodes occupied by each domain. For a fixed rank layout, the scheduler uses that layout, and if the template permits rank remapping, it also specifies the candidates and the rounds for which the mapping holds.
\item \textbf{Communication template.} The template specifies the common round count $V$, the legal steps, and their circuit demands; for a homogeneous template these are $B_v$ and $\mu_p$.
\end{enumerate}

\noindent\textbf{Objective and constraints.}
We first hold the total number of rounds fixed at $V$ and enforce the uplink budget $H$ every round; with one uplink per GPU, this is just the GPU count on the supernode. A configuration interval is a contiguous run of rounds sharing one configuration, so minimizing the number of intervals $K$ is equivalent to minimizing the $K-1$ reconfigurations. Configurations may recur, as in A, B, A, which spans three intervals and two changes. Section~\ref{sec:switch-speed} later allows extra rounds to study the waiting-versus-switching tradeoff.

\noindent\textbf{Outputs.} A solved mega CCL kernel has two parts.
\begin{enumerate}
\item \textbf{Optical configurations.} Each specifies the inter-supernode circuits for its rounds and fits the budget $H$.
\item \textbf{Per-domain schedule.} The schedule gives each domain's steps over $V$ rounds, including the within-phase order or the rank mapping used throughout, and the template generates the corresponding sends, receives, and reductions.
\end{enumerate}

\noindent\textbf{Remark: handling one-directional CCL kernels.} An OCS circuit is full-duplex and offers only a bidirectional path, not a single direction. Operators that already communicate bidirectionally, such as HD and XOR AlltoAll, place demand that is inherently bidirectional and fits these circuits directly, with no extra handling.

Operators such as ring allreduce, PAT, and additive-offset AlltoAll produce only one-directional traffic at the operator level. A general treatment assigns such demand to a bidirectional circuit and counts the idle reverse direction as occupied, which gives an upper bound on the circuit demand and is feasible but wastes bandwidth.

For these naturally homogeneous operators, we use a more efficient construction. We run two kernels in opposite directions and bind them into one scheduling unit that shares one set of circuits, where the send of one kernel and the receive of the other fill the two directions together. The $G$ kernels then form $G/2$ units. The scheduling model below keeps the same form, except that the $G$ standing for the kernel count is replaced by the unit count $G/2$. In the paired formulation, each count represents one fixed pair, $B_v$ gives the circuits required by that pair, and the physical port budget remains $H$.

\subsection{Direct Formulation for the Mega CCL Kernel}
\label{sec:raw-formulation}
As a baseline, we build the mega CCL kernel with a direct formulation that keeps every domain's choice in every round as an explicit variable. These choices determine both the circuit demand and whether the current configuration can carry into the next round.

\noindent\textbf{Per-domain assignment.}
Let the binary variable $x_{g,s,v}$ indicate that domain $g$ executes pattern $v$ in round $s$. Each domain executes exactly one pattern per round and completes its required multiplicities in each phase:
\begin{equation}
\begin{aligned}
\sum_{v\in\mathcal V}x_{g,s,v}&=1, &&\forall g,s,\\
\sum_{s\in I_p}x_{g,s,v}&=\mu_p(v), &&\forall g,p,v,\\
x_{g,s,v}&\in\{0,1\}. &&
\end{aligned}
\label{eq:direct-assignment}
\end{equation}
For $s\in I_p$, \eqref{eq:direct-assignment} assigns each domain a permutation in $\Omega_p$. Property 1 of Definition~\ref{def:homogeneous} makes the concatenated phase permutations legal. Summing $\sum_v x_{g,s,v} B_v$ over all domains gives the aggregate demand in round $s$:
\begin{equation}
D_s(u,y)=\sum_{g=1}^{G}\sum_{v\in\mathcal V}x_{g,s,v}B_v(u,y),\qquad s=1,\ldots,V.
\label{eq:direct-traces}
\end{equation}
The pattern matrices are shared by all domains, so capacity can be checked on each supernode pair after summation.

\noindent\textbf{Circuit capacity and configuration changes.}
Let $Q_s(u,y)$ be the number of circuits between supernodes $u$ and $y$ in round $s$. The matrix is symmetric, nonnegative, and integer, with zero diagonal. Only entries for $u<y$ need variables. The configuration must cover all domain demands:
\begin{equation}
\begin{aligned}
Q_s(u,y)&\ge D_s(u,y), &&\forall s,\ u<y,\\
\sum_{y\ne u}Q_s(u,y)&\le H,&&\forall s,u.
\end{aligned}
\label{eq:direct-capacity}
\end{equation}
A configuration may keep unused circuits within the port budget. When demand saturates every supernode in a round, coverage holds with equality on every edge.

A binary variable $z_s$ records whether the OCS changes configuration between rounds $s$ and $s+1$. We impose
\begin{equation}
\begin{aligned}
-Hz_s&\le Q_{s+1}(u,y)-Q_s(u,y)\le Hz_s,\\
&\forall s<V,\ u<y.
\end{aligned}
\label{eq:direct-switch}
\end{equation}
If $z_s=0$, the configurations must agree. A change in any circuit multiplicity forces $z_s=1$. The direct model is
\begin{equation}
\begin{aligned}
\min_{x,Q,z}\quad &1+\sum_{s=1}^{V-1}z_s,\\
\text{s.t.}\quad &\eqref{eq:direct-assignment},\ \eqref{eq:direct-traces},\\
&\eqref{eq:direct-capacity},\ \eqref{eq:direct-switch},\\
&Q_s(u,y)\in\mathbb Z_{\ge0},\quad z_s\in\{0,1\}.
\end{aligned}
\label{eq:direct-ip}
\end{equation}
The objective counts configuration intervals. Local port reassignment allows the model to record only circuit multiplicities between supernodes. A configuration may cross a phase boundary if it covers the demands on both sides.

\subsection{Two Challenges}
\label{sec:challenges}
The direct model already finds the optimal number of configuration intervals for homogeneous collectives, but applying it to large-scale training raises two challenges.

\noindent\textbf{Challenge 1: fitting non-homogeneous collectives into the model.}
Both the problem definition and the direct model assume the homogeneous representation in Definition~\ref{def:homogeneous}. The upper rows of Table~\ref{tab:natural} list templates that already have both properties, while AlltoAllv and the binary tree in the lower rows must first be put into a common representation and then checked for legality, and their conversion can also change the number of rounds or ports. Table~\ref{tab:natural} also lists the scheduling choice available in each template.

\noindent\textbf{Challenge 2: solving the direct model efficiently.}
Even the direct expansion of a homogeneous template has $G \times V \times |\mathcal V|$ assignment variables, together with $V\binom{M}{2}$ circuit variables and $V-1$ switch indicators. For $G=M=64$, $N=1$, and $V=|\mathcal V|=63$, it uses $254\,016$ assignment variables alone. More importantly, these variables carry a great deal of symmetric redundancy: the number of intervals $K$ depends only on how many domains run each pattern in each round, and permuting the domain labels gives an equivalent solution with the same objective, yet the model still distinguishes the domains explicitly. Reducing the solve size requires compressing the per-domain decisions into compact count variables.

\begin{table}[t]
\centering
\caption{Scheduling choices for the communication templates.}
\label{tab:natural}
\footnotesize
\setlength{\tabcolsep}{3pt}
\renewcommand{\arraystretch}{1.15}
\begin{tabular}{@{}ll@{}}
\toprule
Template & Scheduling choice\\
\midrule
Pairwise AlltoAll & Matching order\\
Fixed ring & Fixed neighbors\\
HD, recursive doubling & Dimension order\\
PAT\textsuperscript{*} & Offset order\\
\midrule
Homogenized AlltoAllv & Matching or offset order\\
Homogenized binary tree & Order at each height\\
\bottomrule
\end{tabular}

\smallskip
\parbox{\columnwidth}{\footnotesize The upper rows are homogeneous as given. *For PAT we use the full-offset template at power-of-two sizes. A fixed pair of reverse domains is one unit.}
\end{table}

\section{Homogenizing Communication Operators}
\label{sec:operator-deployment}

The homogeneous template in Definition~\ref{def:homogeneous} requires a common list, a set of patterns shared by every kernel, where each pattern fixes the peers and circuits of one step. Not every collective naturally has this form. Homogenization converts a collective into such a list while leaving the communication result unchanged, and it must overcome two obstacles.

The first is traffic imbalance. A GPU may exchange unequal amounts with different peers, including zero, so the steps cannot be aligned to common windows without preparation. The second is hierarchical dependency. Some steps have a strict ordering, and a native step may occupy several peers on one GPU at once, so it cannot be treated directly as a separate matching.

We address the two obstacles in turn, using AlltoAllv as the representative case of traffic imbalance and the layered binary tree as the representative case of hierarchical dependency. The padding, fragmentation, and extra rounds introduced by either conversion are all charged to the homogenization cost.

\subsection{Resolving Traffic Imbalance: AlltoAllv}

AlltoAllv is a dynamic exchange primitive common in expert parallelism. The amount each GPU sends to each peer is set by expert routing and may change at every iteration, with no fixed global pairing and with hot peers or zero traffic; one GPU may also send to several peers at once, so a single matching cannot cover all the source-destination demands directly. The point-to-point messages carry no data dependencies, so once the metadata are available their order can be changed, while the source, destination, and size of every message stay fixed.

Before decomposition, one may first rebalance traffic inside a supernode. The scale-up network within a supernode is far faster than the inter-supernode optical paths and uses no optical ports. Over this network, the sender spreads the real traffic of a heavily loaded GPU to other, lightly loaded GPUs in the same supernode, which belong to other domains when $N=1$, so the amount each GPU sends toward every destination supernode is balanced as much as possible; the receiver decouples reaching the correct supernode from reaching the correct GPU, first sending data to a peer GPU at the same local index and then relying on intra-supernode redistribution to forward it to the true destination GPU. This step only rearranges real data and produces no padding. It trades some scale-up time for less padding and fewer reserved windows later, and it can be skipped when the traffic is already balanced.

The traffic is then decomposed into patterns, each fixing the peers and circuits of one GPU at that step. There are three decomposition strategies, with different port, window, and fragmentation costs.

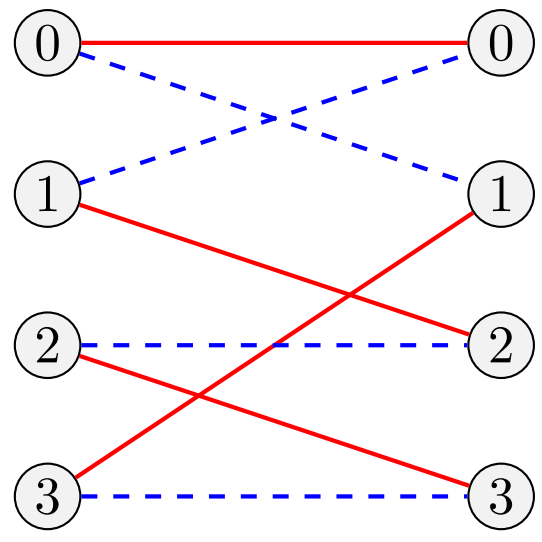
\begin{figure}[t]
\centering
\begin{tikzpicture}[
  >=Stealth,
  lnode/.style={circle,draw,inner sep=1.5pt,fill=gray!10},
  rnode/.style={circle,draw,inner sep=1.5pt,fill=gray!10},
  ma/.style={red,thick},
  mb/.style={blue,thick,dashed}]
  \foreach \i in {0,...,3}{
    \node[lnode] (l\i) at (0,{1.5-\i}) {$\i$};
    \node[rnode] (r\i) at (3,{1.5-\i}) {$\i$};
  }
  \draw[ma] (l0)--(r0); \draw[ma] (l1)--(r2);
  \draw[ma] (l2)--(r3); \draw[ma] (l3)--(r1);
  \draw[mb] (l0)--(r1); \draw[mb] (l1)--(r0);
  \draw[mb] (l2)--(r2); \draw[mb] (l3)--(r3);
\end{tikzpicture}
\caption{Two directed permutations in a BvN decomposition of a $4\times4$ doubly stochastic matrix. Each permutation (solid red or dashed blue) assigns one destination and one source to every GPU.}
\label{fig:bvn-example}
\end{figure}

\noindent\textbf{BvN decomposition (directed permutations, often two ports).}
First round the real byte counts up to window units, then pad the traffic matrix into a nonnegative matrix with equal row and column sums, which is doubly stochastic after normalization. A Birkhoff--von Neumann (BvN) decomposition splits it into directed permutations (Fig.~\ref{fig:bvn-example}): in one permutation every GPU has exactly one destination and one source. Within each window, every GPU has at most one active destination and one active source; when the two differ, two ports are needed. Independent decompositions across kernels need not give the same patterns or multiplicities. For joint scheduling, take the entrywise maximum demand to build a common capacity matrix, pad it until all rows and columns have the same total, and run one BvN decomposition on that padded matrix. Each kernel then fills its real traffic into the allocated windows and leaves unused capacity idle.

\noindent\textbf{Symmetric two-factor decomposition (cycles, two ports).}
First cover the real demands of all kernels with a common bidirectional capacity, then pad it into an undirected graph where every node has equal even degree. Orient the cycles and split the send and receive sides into bipartite perfect matchings, then merge the two copies of each vertex back into a GPU; each pattern is a collection of cycles, a two-factor, where every GPU has two incident edges, counting multiplicity, and uses two ports. To move to one port, these cycles must be split further into matching windows with one peer per GPU and the round count recomputed; an odd cycle requires at least three such windows. The decomposition and port accounting are in our technical report.

\noindent\textbf{Fixed one-factor skeleton (byte-based window reservations, one port).}
When $M\times N$ is even, fix a one-factorization of the complete GPU graph. Let $d^{(g)}_{ij}$ be the real bytes from rank $i$ to rank $j$ in kernel $g$, and let $b$ be the per-direction capacity of a full-duplex circuit in one window. Reserve the following number of common windows for each one-factor pattern $v$:
\begin{equation}
\mu(v)=\max_{\substack{g,\ i<j,\ E_v(\{i,j\})>0}}
\left\{\left\lceil d^{(g)}_{ij}/b\right\rceil,
       \left\lceil d^{(g)}_{ji}/b\right\rceil\right\}.
\label{eq:a2av-window}
\end{equation}
The real data of every source-destination pair are sent only in the windows of its pattern, every kernel reserves circuits with the common multiplicity, and shorter messages leave idle capacity. With $N=1$ and even $M$, each supernode pair belongs to exactly one pattern.

The three strategies are chosen by traffic skew, port budget, and window cost. The schedule assumes the traffic characteristics are known before compilation, with message volumes and multiplicities obtained from offline measurement or prediction, and windows are reserved accordingly; only when real traffic exceeds the reserved capacity is the template updated and recompiled. These homogenizations all incur preprocessing cost, and padding and window alignment reserve some idle ports or bandwidth; if homogenization adds rounds, the extra rounds are charged to the conversion cost and later scheduling uses the new round count.

\subsection{Resolving Hierarchical Dependency: The Binary Tree}

The second obstacle is the hierarchical dependency between steps, for which the layered binary-tree AllReduce is the representative case. Binary trees are commonly used for small-message, low-latency AllReduce, with all GPUs forming a fixed rooted tree and the operation split into a bottom-up reduction and a top-down broadcast. It has two features that block a direct homogeneous template: strict inter-level dependency, where a child must finish before its parent, and an internal node that may communicate with several children at once, so a native step occupies several peers and is not a one-peer-per-GPU matching. Homogenization rearranges the tree into one-port matchings while keeping the inter-level dependencies.

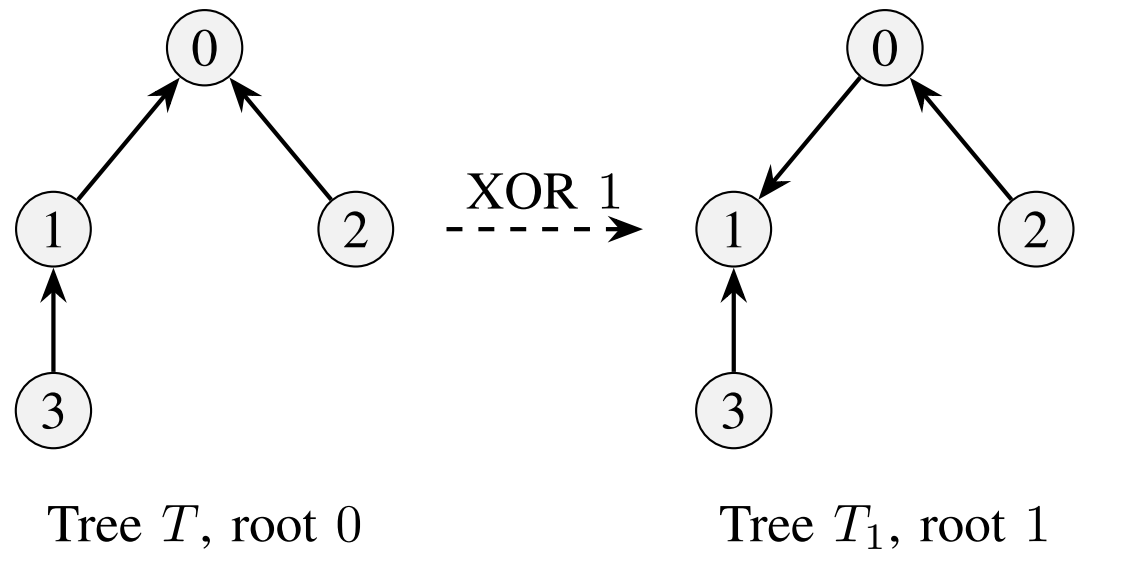
\begin{figure}[t]
\centering
\begin{tikzpicture}[
  >=Stealth,
  tnode/.style={circle,draw,inner sep=2pt,fill=gray!10},
  tedge/.style={->,>=Stealth,thick}]
  \begin{scope}[xshift=0cm]
    \node[tnode] (a0) at (0,1.2) {0};
    \node[tnode] (a1) at (-1,0) {1};
    \node[tnode] (a2) at (1,0) {2};
    \node[tnode] (a3) at (-1,-1.2) {3};
    \draw[tedge] (a1)--(a0); \draw[tedge] (a2)--(a0); \draw[tedge] (a3)--(a1);
    \node[below] at (0,-1.7) {Tree $T$, root $0$};
  \end{scope}
  \begin{scope}[xshift=4.5cm]
    \node[tnode] (b0) at (0,1.2) {0};
    \node[tnode] (b1) at (-1,0) {1};
    \node[tnode] (b2) at (1,0) {2};
    \node[tnode] (b3) at (-1,-1.2) {3};
    \draw[tedge] (b0)--(b1); \draw[tedge] (b3)--(b1); \draw[tedge] (b2)--(b0);
    \node[below] at (0,-1.7) {Tree $T_1$, root $1$};
  \end{scope}
  \draw[->,thick,dashed] (1.6,0) -- node[above]{XOR $1$} (2.9,0);
\end{tikzpicture}
\caption{XOR translation for one block. With $M=4$, translating all tree edges by 1 moves the root from 0 to 1. The physical nodes stay in place, and the tree retains its shape and dependencies.}
\label{fig:tree-reroot}
\end{figure}

First split each GPU's input of size $D$ into $M$ equal blocks of size $D/M$ (the first step of the framework), each keeping the full hierarchical dependency while the blocks are mutually independent. Let $M=2^L$ and $N=1$, let $T$ be the directed edge set of the original binary tree rooted at 0, with each edge denoting the reduction dependency from child to parent. For the data block indexed $g$, translate every node label of the original tree by XOR, giving its translated tree
\begin{equation}
T_g=\{(u\oplus g)\to(v\oplus g):(u\to v)\in T\},
\label{eq:tree-shift}
\end{equation}
which has the same tree shape and hierarchical dependency as the original, with the root moved to $g$ (Fig.~\ref{fig:tree-reroot}). For any edge $u\to v$ of the original tree, execute all its translated copies in parallel: GPU $x$ sends block $x\oplus u$ to $x\oplus u\oplus v$ and receives block $x\oplus v$ from that peer. Every GPU connects one peer, with one $D/M$ block in each direction, globally forming an undirected perfect matching of offset $u\oplus v$.

Phases are grouped by the subtree height of the child on each edge: edges at one height have no ancestral dependency and can be reordered freely within the phase, and before a block sends to its parent it has already collected all subtree contributions. After reduction, broadcast follows the heights in reverse with each edge reversed, and the order within each height remains independently selectable. Reduction and broadcast each use $M-1$ matching rounds and send $(M-1)D$ bytes across the network in total, as in the unsplit tree; compared with native level-parallel execution, the round count may increase.

A double binary tree (DBT) uses two complementary mirror trees, each handling half of the tensor. Its key difference from the homogenized operators above is that once a kernel selects a rank mapping at the first level, the locations of intermediate results are fixed and every level of reduction and broadcast must keep that mapping, so it cannot be counted round by round but must be counted over the complete mapping. The count model, correctness proof, and closed-form construction for this case are in our technical report.

\section{Count Model for the Mega CCL Kernel}
\label{sec:method}

The direct formulation in Section~\ref{sec:raw-formulation} solves the mega CCL kernel with a separate choice for each domain through the per-domain variable $x_{g,s,v}$. Swapping the complete schedules of two domains leaves the aggregate circuit demand in every round unchanged, since both domains use the same homogeneous template. We therefore replace the per-domain variables with pattern counts $C_s(v)$, which record how many domains execute pattern $v$ in round $s$, and solve the same mega CCL kernel in this compact count form. This section proves that the counts recover a legal schedule for every domain and identifies when the variables can be compressed further.

\subsection{From a Schedule to Pattern Counts}
\label{sec:count-table}
We first sum the columns of the four-domain schedule.

\noindent\textbf{Counting the example.}
Recall the four-kernel example in Table~\ref{tab:toy}, with rounds $s\in\{1,2,3\}$ and pattern set $\mathcal V=\{a,b,c\}$. For this example, let $C_s(v)$ denote the number of kernels that execute pattern $v$ in round $s$. The count row of the table gives
\begin{equation}
\begin{aligned}
&\text{Round 1} && C_1(a)=2,\ C_1(b)=1,\ C_1(c)=1,\\
&\text{Round 2} && C_2(a)=2,\ C_2(b)=1,\ C_2(c)=1,\\
&\text{Round 3} && C_3(a)=0,\ C_3(b)=2,\ C_3(c)=2.
\end{aligned}
\end{equation}
Rounds 1 and 2 have the same counts, $2a+b+c$, and therefore the same circuit demand, while Round 3 has $2b+2c$. Since a count vector fixes the circuit demand in a homogeneous template, the first two rounds share one configuration and only the third round needs reconfiguration.

\noindent\textbf{General definition.}
Let $a_{g,s}$ be the pattern that domain $g$ executes in round $s$, so $x_{g,s,v}=\mathbf1[a_{g,s}=v]$. The pattern count is
\begin{equation}
C_s(v)=\sum_{g=1}^{G}x_{g,s,v}
=\sum_{g=1}^{G}\mathbf1[a_{g,s}=v],
\label{eq:projection}
\end{equation}
This linear projection keeps the number of domains using each pattern and discards their identities. Several assignments can have the same count table. Since each domain uses exactly one pattern per round,
\begin{equation}
\sum_{v\in\mathcal V}C_s(v)=G,\qquad C_s(v)\in\mathbb Z_{\ge0}.
\label{eq:phi}
\end{equation}
Each domain contains $M \times N$ GPUs, but they jointly execute one step, so a domain contributes one count in a round regardless of $N$, and the counts in a round sum to $G$.

The counts must also match how many times each domain runs every pattern. Across the whole call, each domain runs pattern $v$ a total of $\mu(v)$ times, and within phase $p$ it runs $v$ a total of $\mu_p(v)$ times. Summing over the $G$ domains gives
\begin{equation}
\sum_{s=1}^V C_s(v)=G\mu(v),\qquad
\sum_{s\in I_p}C_s(v)=G\mu_p(v).
\label{eq:physical-count}
\end{equation}
The first equation fixes the total count of pattern $v$ over all rounds, and the second fixes its count within each phase.

\subsection{Integer Formulation for a Homogeneous Template}
\label{sec:count-ip}
We now state the scheduling problem as an integer program. It partitions the $V$ rounds into the fewest contiguous configuration intervals and decides the pattern count $C_s(v)$ in every round, subject to a small set of linear constraints.

\noindent\textbf{Interval partition.}
A configuration interval is a contiguous run of rounds that share one optical configuration. Let $w_k$ be the number of rounds in interval $k$, a positive integer. The $K$ intervals cover the rounds in order without skipping or adding any round, so their widths sum to $V$.

\noindent\textbf{Round counts.}
Let $J_k$ be the set of rounds in interval $k$. The counts $C_s(v)$ are nonnegative integers that sum to $G$ in every round. They need not be equal across the rounds inside one interval, since the interval only requires one shared configuration to cover every round's demand.

\noindent\textbf{Phase totals.}
Equation~\eqref{eq:physical-count} fixes how many times each pattern must be used within every phase. An interval may span a phase boundary and hold one configuration across two phases, but each round still counts toward the phase it belongs to.

\noindent\textbf{Circuit capacity.}
The aggregate demand in round $s$ is $D_s=\sum_v C_s(v)B_v$. The configuration $Q_k$ must provide at least $D_s$ on every edge for each round in $J_k$, and the number of circuits incident to every supernode must stay within the port budget $H$.

The integer formulation is
\begin{equation}
\begin{aligned}
\text{Objective}\quad &\min K,\\
\text{Partition}\quad &\sum_{k=1}^K w_k=V,\quad w_k\in\mathbb Z_{>0},\\
\text{Round counts}\quad &\sum_{v\in\mathcal V}C_s(v)=G,\quad C_s(v)\in\mathbb Z_{\ge0},\\
\text{Phase totals}\quad &\sum_{s\in I_p}C_s(v)=G\mu_p(v),\quad \forall p,v,\\
\text{Capacity}\quad &Q_k(u,y)\ge D_s(u,y),\quad s\in J_k,\ u<y,\\
&\sum_{y\ne u}Q_k(u,y)\le H,\quad \forall k,u.
\end{aligned}
\label{eq:ip-model}
\end{equation}

An equivalent implementation uses one configuration per round and a binary indicator at each boundary, as in the direct model, and minimizes the number of changes as a linear integer program. Both forms describe the same problem, and our technical report gives the details.

\subsection{Recovering a Schedule for Each Domain}
\label{sec:method-recover}
The count model decides only the pattern counts $C_s(v)$, but a mega CCL kernel must ultimately specify the per-domain variables $x_{g,s,v}$, which tell every domain what to do. The input $\mu_p(v)$ states how many times each domain must run every pattern, and the counts place these executions across the rounds. Although the counts no longer say which domain does what, the theorem below proves that they can be turned back into a legal $x_{g,s,v}$ for every domain.

\begin{theorem}[Exact recovery from pattern counts]
\label{th:exact}
Under Definition~\ref{def:homogeneous}, for any $N$, every nonnegative integer table $C$ that satisfies the round totals in \eqref{eq:phi} and the phase totals in \eqref{eq:physical-count} admits a legal per-domain assignment $x_{g,s,v}$. Moreover, the count model~\eqref{eq:ip-model} and the direct model~\eqref{eq:direct-ip} admit the same configuration sequences and attain the same optimum.
\end{theorem}
\begin{proof}
The forward direction is the projection itself, since summing $x$ over domains gives $C$. For the reverse, fix a phase $p$ and build a bipartite multigraph. Place the $|I_p|$ rounds on the left, and on the right place $\mu_p(v)$ copies of each pattern $v$. Each round contains $G$ pattern occurrences, and each copy is filled with $G$ occurrences of its pattern. Connect every occurrence by an edge from its round on the left to its copy on the right, allowing parallel edges.

By the two conservation laws, every vertex has degree $G$. Take any set $S$ of left vertices. Its $G|S|$ edges lead to neighbors of degree $G$, so those neighbors number at least $|S|$, that is, $|N(S)|\ge|S|$. The two sides have equal size, so Hall's theorem~\cite{hall-representatives}, the bipartite form of the Birkhoff-von Neumann decomposition of demand matrices~\cite{birkhoffvn,birkhoffrevisit}, yields a perfect matching. Remove this matching and every vertex degree drops by one; repeating this $G$ times decomposes the graph into $G$ perfect matchings. Assign one matching to each domain. Each matching gives a domain one task in every round and exactly the required copies of every pattern in this phase. Concatenating the assignments across phases gives a full schedule, which Property 1 of Definition~\ref{def:homogeneous} guarantees to be legal. In every round
\begin{equation}
\sum_g x_{g,s,v}=C_s(v),\qquad D_s=\sum_v C_s(v)B_v.
\label{eq:surjection}
\end{equation}
Recovery keeps the aggregate demand unchanged, so every configuration in the count solution stays feasible, and the first identity confirms that summing the recovered $x$ returns the same $C$. The two directions therefore preserve the same configuration sequences, and the count model reaches exactly the same optimum as the direct model.
\end{proof}

The assignment produced by the matching construction need not be unique: when more than one legal assignment realizes the same counts, the construction returns one of them rather than reconstructing the original per-domain table.

\subsection{Further Reductions}
\label{sec:readback}

The sections above compress the per-domain variables $x_{g,s,v}$ into the pattern counts $C_s(v)$ without changing the optimum. This section reduces the variables further, using the rule that maps a count vector to its circuit demand. It covers when rounds can share a configuration, when one interval needs only a single count vector, and when several patterns share one resource matrix.

\noindent\textbf{When rounds can share a configuration.}
First write the rule that turns a count vector into circuit demand,
\begin{equation}
\Phi(C)(u,y)=\sum_v C(v)B_v(u,y).
\label{eq:interval-count}
\end{equation}
For a contiguous set $J$ of rounds, one configuration covers all of them exactly when it meets the port budget after taking, on each edge, the largest demand over these rounds:
\begin{equation}
Q_J(u,y)=\max_{s\in J}\Phi(C_s)(u,y),\qquad
\sum_{y\ne u}Q_J(u,y)\le H,\quad \forall u.
\label{eq:cover}
\end{equation}
The port inequality is therefore both necessary and sufficient, and the counts may still differ across the rounds. Given a fixed count sequence, repeatedly extending the longest feasible prefix in this way produces the fewest intervals. Our technical report proves this.

\noindent\textbf{One count vector per interval.}
Now suppose a feasible configuration can cover only one count vector, so that no configuration within the port budget covers two distinct feasible vectors. The configuration then determines its count vector, and all rounds in one interval share a single vector $C_k$ that replaces the per-round variables.
A sufficient condition is that every supernode is saturated and that $\Phi$ is injective on the feasible vectors. Injectivity only needs to be checked on feasible differences $d=C-C'$:
\begin{equation}
\sum_v d(v)B_v=\mathbf 0\ \Longrightarrow\ d=0.
\end{equation}
Let $w_{kp}=|J_k\cap I_p|$. Under this condition, phase conservation becomes $\sum_k w_{kp}C_k(v)=G\mu_p(v)$ and we can set $Q_k=\Phi(C_k)$. When the condition fails, forcing one constant count over an interval can exclude legal schedules, and the per-round model~\eqref{eq:ip-model} remains valid. Our technical report verifies the condition for common templates.

\noindent\textbf{Grouping identical resource matrices.}
Different GPU patterns can lead to the same inter-supernode demand, that is, the same matrix $B_v$. Group every such pattern into a class $h$ and count the class directly,
\begin{equation}
C_s^{\rm cls}(h)=\sum_{v\in h}C_s(v),
\label{eq:class}
\end{equation}
For each domain, first recover the order of classes, then assign the class positions to concrete patterns according to their multiplicities within the phase, and local port reassignment carries out the choice. This reduction keeps the optimum for every homogeneous template, even when one configuration admits several count vectors. Our technical report gives a power-of-two XOR example.

\subsection{Constructions for Large Instances}
\label{sec:size}
For very large instances, solving the integer program can be expensive. The four constructions below, cyclic grouping, pattern blocking, domain replication, and domain overlay, build feasible schedules directly and upper-bound the number of intervals. If the exact optimum is needed, the count model above still applies.

\noindent\textbf{Cyclic grouping.}
Consider a reorderable phase in which each pattern uses at most $r$ ports per supernode and $H\ge Gr$. Suppose $d$ patterns must each run once and $d$ divides $G$. Split the $G$ domains into $d$ groups of $G/d$ domains, and rotate the groups across the patterns over $d$ rounds. In every round each pattern carries $G/d$ domains, and every domain visits all $d$ patterns. Since every round has the same aggregate demand, one configuration covers all $d$ rounds.

When $G=2^h$, divide the $|I_p|$ pattern occurrences in this phase into groups of $G$, where each full group needs one interval. Write the remaining number in binary and form one group for each bit that is set, so every group size divides $G$. This gives
\begin{equation}
K_{\rm cycle}\le
\left\lfloor\frac{|I_p|}{G}\right\rfloor+
\operatorname{popcount}(|I_p|\bmod G).
\label{eq:cycle}
\end{equation}
Apply the construction to each phase on its own, then check whether the configurations at phase boundaries can be merged.

\noindent\textbf{Pattern blocking.}
Partition a phase into several blocks and have every domain follow the same block order. Concatenating the per-block schedules gives an interval count bounded by the sum over blocks, $K_{\rm out}\le\sum_\ell K^{(\ell)}$, at the cost of forbidding reordering across blocks. Turning this into an additive error bound in terms of the block boundaries needs two conditions, an exact solution for every block and the restricted feasibility condition given in our technical report.

\noindent\textbf{Domain replication.}
Take a schedule that serves $g_0$ domains under budget $H_0$ and replicate it $t$ times. The replicated schedule serves $tg_0$ domains under budget $tH_0$ with exactly the same interval boundaries, so
\begin{equation}
K^*(tg_0,tH_0)\le K^*(g_0,H_0).
\label{eq:copy}
\end{equation}
Both sides use the same task template, hence the inequality is a feasible upper bound.

\noindent\textbf{Domain overlay.}
Two independent groups with budgets $H_1$ and $H_2$ can share one round axis. Split the axis at the union of the two groups' boundaries and add their configurations on every resulting interval; the combined budget is $H_1+H_2$. If the two schedules have $K_1$ and $K_2$ intervals and share $r$ internal boundaries, the refinement has
\begin{equation}
K_{\rm overlay}=K_1+K_2-1-r,
\label{eq:overlay}
\end{equation}
Merging adjacent identical configurations can lower the count further, and running the overlay in real time also requires the round durations to align.

Replication keeps the interval count unchanged whenever the domain count and the budget grow by the same factor, and the overlay bound is tighter when more boundaries coincide. Our technical report states the applicability conditions, error bounds, and small-instance checks.

\section{Evaluation}
\label{sec:evaluation}

We evaluate four aspects: how much letting domains take independent legal step orders cuts circuit reconfigurations, how those fewer reconfigurations improve end-to-end training efficiency, the tradeoff between reconfiguration cost and execution rounds when the round count is relaxed, and how much pattern counts reduce offline solving time. The interval counts and their optimality certificates come from offline solvers, and the completion times come from ASTRA-sim training replays.

\subsection{Reconfiguration Counts}

At the largest fully loaded point, $M=G=H=64$, XOR AlltoAll takes 63 rounds: without scheduling each round requests a different matching, giving 62 reconfigurations, while joint scheduling needs only 3. HD AllReduce takes 12 rounds and goes from 10 reconfigurations to 2; its last reduce-scatter and first all-gather round already share one connection.

\begin{table}[h!]
\centering
\caption{Mean number of configuration intervals for three collectives at each port budget. Each instance has one fewer reconfiguration than intervals. Our technical report gives the full results.}
\label{tab:exact}
\setlength{\tabcolsep}{3.5pt}
\begin{tabular}{llrr}
\toprule
Collective & $H$ & Unscheduled & Scheduled\\
\midrule
AlltoAll & 4  & 2.000  & 1.333\\
AlltoAll & 8  & 5.000  & 2.000\\
AlltoAll & 16 & 11.000 & 2.333\\
AlltoAll & 32 & 23.000 & 2.765\\
AlltoAll & 64 & 47.000 & 3.030\\
\midrule
AlltoAllv & 4  & 3.889  & 1.222\\
AlltoAllv & 8  & 9.400  & 1.867\\
AlltoAllv & 16 & 19.926 & 2.370\\
AlltoAllv & 32 & 42.961 & 2.706\\
AlltoAllv & 64 & 88.465 & 2.919\\
\midrule
HD & 4  & 2.000  & 1.000\\
HD & 8  & 4.000  & 2.000\\
HD & 16 & 6.000  & 2.000\\
HD & 32 & 8.000  & 2.000\\
HD & 64 & 10.000 & 3.000\\
\bottomrule
\end{tabular}
\end{table}

To check whether the gain holds across scales and traffic profiles, we run a grid sweep on the three collectives. Every instance is fully provisioned with one GPU per domain on each supernode, so $N=1$ and $G=H$. The port budget $H$ ranges over 4, 8, 16, 32, and 64, and the number $M$ of supernodes ranges from $H/2$ to $H$, so both even and odd $M$ appear and the ratios of domains to supernodes differ. AlltoAll and HD use one template at each size. For AlltoAllv we draw three deterministic traffic profiles with a fixed seeded sampler, from low to high skew; the profiles differ only in per-peer multiplicities, not in rank placement. This gives $33\times3=99$ AlltoAllv instances at $H=64$. The separate double binary tree uses $H=2G$, as it cannot be counted round by round. Table~\ref{tab:exact} reports the mean configuration intervals without scheduling and with scheduling. Regardless of scale and traffic, scheduling keeps the mean interval count between 1 and 3. Without scheduling the count grows quickly with the port budget, reaching 47 for AlltoAll and 88.465 for AlltoAllv at $H=64$, while scheduling still returns about 3, so the gain grows with scale.

For instances too large to solve exactly, the constructions in Section~\ref{sec:size} produce a feasible interval count directly, without solving the full instance. Table~\ref{tab:construct} compares these constructions with the two exact count models on the same templates: the per-round and per-interval models are the two exact counts, defined in Section~\ref{sec:count-ip} and Section~\ref{sec:readback}, and cyclic grouping, blocking, replication, and overlay are the four constructions. The two exact counts agree on templates where each configuration determines its counts.

\begin{table}[h!]
\centering
\caption{Mean configuration intervals for the constructions in Section~\ref{sec:size} and the two exact count models on the same templates. The exact models agree on templates for which each configuration determines its counts.}
\label{tab:construct}
\setlength{\tabcolsep}{3pt}
\begin{tabular}{llrrrrrr}
\toprule
Collective & $H$ & Round & Interval & Cyclic & Block & Copy & Overlay\\
\midrule
AlltoAll & 4  & 1.333 & 1.333 & 1.333 & 1.333 & 1.333 & 1.667\\
AlltoAll & 8  & 2.000 & 2.000 & 2.000 & 2.000 & 2.000 & 2.800\\
AlltoAll & 16 & 2.333 & 2.333 & 2.556 & 2.333 & 2.556 & 3.556\\
AlltoAll & 32 & 2.765 & 2.765 & 3.059 & 2.824 & 2.882 & 4.412\\
AlltoAll & 64 & 3.030 & 3.030 & 3.545 & 3.273 & 3.273 & 5.091\\
\midrule
AlltoAllv & 4  & 1.222 & 1.222 & 1.333 & 1.333 & 1.333 & 1.667\\
AlltoAllv & 8  & 1.867 & 1.867 & 2.333 & 2.267 & 2.667 & 3.267\\
AlltoAllv & 16 & 2.370 & 2.370 & 3.074 & 2.889 & 4.074 & 5.444\\
AlltoAllv & 32 & 2.706 & 2.706 & 3.471 & 3.333 & 6.216 & 7.275\\
AlltoAllv & 64 & 2.919 & 2.919 & 4.071 & 3.949 & 8.394 & 8.222\\
\midrule
HD & 4  & 1.000 & 1.000 & 1.000 & 1.000 & 1.000 & 1.000\\
HD & 8  & 2.000 & 2.000 & 2.500 & 2.500 & 2.500 & 2.500\\
HD & 16 & 2.000 & 2.000 & 2.500 & 2.500 & 2.500 & 2.500\\
HD & 32 & 2.000 & 2.000 & 2.500 & 2.500 & 2.500 & 2.500\\
HD & 64 & 3.000 & 3.000 & 4.000 & 4.000 & 4.000 & 4.000\\
\bottomrule
\end{tabular}
\end{table}

The constructions are feasible upper bounds rather than exact optima: at $H=64$ they average about four intervals against the exact optimum of three over the $M=32,64$ cases, while all four reach the HD optimum at $H=4$.

Every reported optimum is backed by either an integer lower bound or an exhaustive exclusion that rules out smaller counts, together with a certificate realizing the reported count, so it is certified rather than merely feasible. Our technical report gives the procedure and the integer certificates.

\subsection{Effect on Completion Time}

We replay synthetic training workloads in ASTRA-sim using parameters from eight real models. The deployment has $M=G=H=64$, or 4{,}096 logical participants, with 400 Gb/s per port and an OCS reconfiguration delay of $10\,\mu$s. Each layer executes two EP AlltoAllv calls in the forward pass and two in the backward pass. A DP ring collective closes the iteration.

Every EP call in this replay uses the same 63-round pairwise exchange template, and scheduling reduces its reconfigurations from 62 to 3. The DP ring remains static in both schedules. Each EP AlltoAllv is replayed once with uniform traffic and once with skewed traffic. Relative to unscheduled execution, iteration time falls by 8.84\%--34.34\% with uniform traffic and by 4.89\%--20.27\% with skewed traffic.

\subsection{Trading Rounds for Reconfigurations}
\label{sec:switch-speed}

The fixed-round problem requires each domain to execute one task in every round. Allowing a domain to wait may extend the schedule but can keep a configuration in use longer. We study when the saved reconfigurations justify the extra windows, using AlltoAllv as an example.

Suppose a configuration provides $c_v$ parallel copies of a matching. At most $c_v$ domains can use it per round. With $G=4$ and $c_v=2$, four domains need at least two rounds to execute the matching once each. A new configuration can allocate more copies to that matching and clear the queue sooner, at the cost of a reconfiguration.

Each domain has $V$ active steps. Let $T\ge V$ be the elapsed number of rounds, including waiting. Allowing more configuration intervals can only decrease or preserve the minimum round count. Completion time depends on the rounds and the cost of switching.

\noindent\textbf{Minimum rounds with a static configuration.}
Assume $G=H$, one reorderable phase, and edge-disjoint perfect matchings. Every domain executes matching $v$ a total of $\mu_v$ times, with equal chunks, so $V=\sum_v\mu_v$. A configuration with $c_v$ copies of matching $v$ can serve $c_v$ domains on it in one round. One copy uses one port per supernode, hence $\sum_v c_v\le G$.

Each domain needs at least $V$ rounds. Pattern $v$ has $G\mu_v$ tasks and capacity $c_v$ per round. For a configuration that covers every active pattern, the exact number of rounds is
\begin{equation}
T(c)=\max\left\{V,\ \max_v\left\lceil\frac{G\mu_v}{c_v}\right\rceil\right\}.
\label{eq:static-rounds}
\end{equation}
Bipartite edge coloring attains both lower bounds simultaneously. Our technical report gives the proof.

Optimizing the capacities under the port budget gives the static optimum
\begin{equation}
T_1=\min\left\{t\in\mathbb Z:\ t\ge V,\
       \sum_v\left\lceil\frac{G\mu_v}{t}\right\rceil\le G\right\}.
\label{eq:best-static-rounds}
\end{equation}
To finish in $t$ rounds, each pattern needs at least $\lceil G\mu_v/t\rceil$ copies. Every active pattern needs at least one. Thus waiting alone cannot make a static configuration feasible when there are more active patterns than ports. In our sweep, $|\mathcal V|=M-1<G=M$, so a static schedule always exists.

\noindent\textbf{When a reconfiguration saves time.}
We use the same session assumption for optical and electrical transmission. A domain pays startup $\alpha$ once for each distinct peer. It can reuse that session even after communicating with other peers in intervening rounds. Every domain here has $M-1$ active peers, so startup contributes $(M-1)\alpha$ on the critical path. Let $b$ be the chunk size and $B$ the port bandwidth. The initial configuration is ready before communication, each window lasts $b/B$, and switching does not overlap transmission. Completion time is
\begin{equation}
t_{\rm OCS}(T,K;\tau)=T b/B+(M-1)\alpha+(K-1)\tau .
\label{eq:round-switch-time}
\end{equation}
A schedule with $K>1$ intervals finishes earlier than the optimal static schedule of $T_1$ rounds precisely when
\begin{equation}
\tau<\frac{(T_1-T)b}{B(K-1)},\qquad K>1.
\label{eq:round-switch-threshold}
\end{equation}
The two schedules contact the same peers, so their startup costs cancel. The right-hand side is the transmission-window time saved per reconfiguration. With the static schedule included, the optimal number of changes is nonincreasing in $\tau$. For sufficiently large $\tau$, a static schedule attains the minimum time $T_1b/B+(M-1)\alpha$. Our technical report gives the monotonicity proof.

We instantiate this study with $M=G=H\in\{4,8,16,32,64\}$ and one GPU per domain per supernode. Pattern $v$ pairs $u$ with $u\mathbin{\oplus}v$. In the representative workload, $\mu_v=\max\{1,\lfloor M/v+1/2\rfloor\}$. All domains use the same matrix. Peer volumes vary, but every supernode sends and receives the same total volume. The synthetic workload requires no receiver hotspot or free redistribution. For mean peer message size $P$, set $b=(M-1)P/V$ so each GPU sends $(M-1)P$ bytes under every schedule. Bandwidth $B=225$ GB/s and startup $\alpha=0.5\,\mu$s are timing parameters.

Equation~\eqref{eq:best-static-rounds} gives the optimal static configuration. For dynamic candidates, we sort patterns by decreasing workload, partition that list into contiguous groups, and compute each group's minimum round count using \eqref{eq:static-rounds} with its allocated capacities optimized. Dynamic programming selects the group boundaries. This gives recoverable $(T,K)$ candidates within this ordered, disjoint-group family. Allowing a pattern to span intervals can improve the result. Our technical report gives a counterexample.

At 64 supernodes, each domain has 300 active steps. A static schedule needs 2{,}048 rounds, one reconfiguration reduces this to 338, four intervals give 305 rounds, and nine intervals reach the work bound of 300. Going from 338 to 300 needs seven more changes, so the schedule with the fewest rounds need not finish first. At $P=100$ KB, the 338-round schedule is faster than the static one for $\tau<159.60\,\mu$s, while at $\tau\ge163.15\,\mu$s even the work bound cannot justify one change, and these two values bracket the break-even point.

\subsection{Offline Planning Time}

This section measures the offline planning time of the count model and compares it with the direct formulation. We implement two count variants: the per-round count keeps one count vector for every round and applies to any template, as defined in Section~\ref{sec:count-ip}, and the per-interval count lets all rounds in one interval share a vector when a configuration determines its counts, with smaller variables and certificates, as defined in Section~\ref{sec:readback}. Table~\ref{tab:overhead} reports the mean planning time of the two variants on the same machine and fully loaded instances, together with the time of the four constructions in Section~\ref{sec:size}, cyclic grouping, blocking, replication, and overlay, which cover different work from the exact counts.

\begin{table}[h!]
\centering
\caption{Mean planning time (ms) on the same machine and fully loaded instances. For the two count models, timings include startup, construction, search, and certificate replay. The construction columns time only the construction calls and cover different work from the solvers; values below one microsecond show as 0.000.}
\label{tab:overhead}
\setlength{\tabcolsep}{2.4pt}
\begin{tabular}{llrrrrrr}
\toprule
Collective & $H$ & Round & Interval & Cyclic & Block & Copy & Overlay\\
\midrule
AlltoAll & 4  & 5.496 & 4.933 & 0.000 & 0.017 & 0.012 & 0.022\\
AlltoAll & 8  & 5.248 & 5.162 & 0.000 & 0.019 & 0.018 & 0.025\\
AlltoAll & 16 & 6.058 & 5.724 & 0.000 & 0.030 & 0.021 & 0.043\\
AlltoAll & 32 & 9.633 & 8.816 & 0.000 & 0.069 & 0.032 & 0.066\\
AlltoAll & 64 & 85.254 & 84.051 & 0.000 & 0.181 & 0.047 & 0.115\\
\midrule
AlltoAllv & 4  & 5.201 & 4.881 & 0.000 & 0.013 & 0.009 & 0.017\\
AlltoAllv & 8  & 5.434 & 5.022 & 0.000 & 0.022 & 0.016 & 0.027\\
AlltoAllv & 16 & 6.184 & 5.862 & 0.000 & 0.050 & 0.030 & 0.050\\
AlltoAllv & 32 & 9.900 & 9.024 & 0.000 & 0.144 & 0.046 & 0.074\\
AlltoAllv & 64 & 44.910 & 44.895 & 0.000 & 0.529 & 0.085 & 0.150\\
\midrule
HD & 4  & 5.158 & 4.915 & 0.000 & 0.021 & 0.015 & 0.020\\
HD & 8  & 5.573 & 4.938 & 0.000 & 0.020 & 0.020 & 0.028\\
HD & 16 & 5.718 & 5.246 & 0.000 & 0.041 & 0.018 & 0.034\\
HD & 32 & 6.399 & 6.117 & 0.000 & 0.038 & 0.039 & 0.069\\
HD & 64 & 11.241 & 10.868 & 0.000 & 0.061 & 0.043 & 0.109\\
\bottomrule
\end{tabular}
\end{table}

\noindent\textbf{Direct formulation versus per-round counts.}
On the hardest skewed AlltoAllv instance at $H=8$, the direct model takes 29\,009 seconds (nearly eight hours) to prove $K=3$, while the per-round count DP solves the same instance in milliseconds.

\noindent\textbf{Per-round versus per-interval counts.}
When a configuration determines its counts, one count vector per interval suffices, and the two dynamic programs have similar running times because both use interval maxima in the search. Interval counts reduce the variable and certificate sizes under the stated condition, while the per-round model remains available outside that scope.

\section{Conclusion}
\label{sec:conclusion}

A large-model deployment runs many CCL kernels on the domains that share one set of optical ports. These kernels do not have to stay in step. When a collective allows its steps to be reordered, different kernels can take different legal steps in the same round, so the same circuits can stay in place across rounds and need far fewer reconfigurations.

We group these interchangeable kernels into one mega CCL kernel that schedules their shared ports together. Instead of deciding every kernel separately, the count model records how many kernels run each pattern in a round, and bipartite matching turns these counts into a legal schedule for each kernel, with the same optimum as the direct formulation. For irregular AlltoAllv and the layered binary tree, we first convert them into a common representation, and the same count and matching step then applies.

On fully loaded deployments, the mega CCL kernel removes most of the reconfigurations, shortens training iteration time, and keeps exact offline solving practical while reaching scales that the direct formulation cannot.

\end{document}